\documentclass{article}

\usepackage{indentfirst}
\usepackage{graphicx}
\usepackage{mdframed}
\usepackage{url}

\begin{document}

\title{On the Importance of Correlations\\in Rational Choice:\\A Case for Non-Nashian Game Theory}

\author{Ghislain Fourny\\ETH Z\"urich\\ghislain.fourny@inf.ethz.ch}

\maketitle

\begin{abstract}
The Nash equilibrium paradigm, and Rational Choice Theory in general, rely on agents acting independently from each other. This note shows how this assumption is crucial in the definition of Rational Choice Theory. It explains how a consistent Alternate Rational Choice Theory, as suggested by Jean-Pierre Dupuy, can be built on the exact opposite assumption, and how it provides a viable account for alternate, actually observed behavior of rational agents that is based on correlations between their decisions. The end goal of this note is three-fold: (i) to motivate that the Perfect Prediction Equilibrium, implementing Dupuy's notion of projected time and previously called ``projected equilibrium'', is a reasonable approach in certain real situations and a meaningful complement to the Nash paradigm, (ii) to summarize common misconceptions about this equilibrium, and (iii) to give a concise motivation for future research on non-Nashian game theory.
\end{abstract}

\section{Dependency in Rational Choice}

[1] Rational Choice Theory (RCT) is at the core of the predominant school of thought in microeconomics: neoclassical economics. From a high-level perspective, it comes down to three assumptions as stated by Weintraub \cite{weintraub}:

\begin{enumerate}
\item People have rational preferences among outcomes.
\item Individuals maximize utility and firms maximize profits.
\item People act independently on the basis of full and relevant information.
\end{enumerate}

\begin{figure}
\begin{mdframed}
\includegraphics[width=\textwidth]{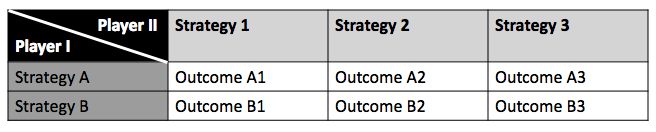}
\caption{A game in normal form with two players and 2 resp. 3 strategies, making 6 outcomes.} 
\label{fig-normal-form}
\end{mdframed}
\end{figure}

[2] The assumption of independence is at the core of the definition of the Nash equilibrium \cite{nash} for games in normal form. A game in normal form, as shown in Figure \ref{fig-normal-form} is expressed---for two players---as a matrix with the rows reprensenting the strategies available to one player, and the columns those available to the other player. In the Nash paradigm, each player holds the opponent's strategy (say, a column) for fixed while maximizing their payoff (say, picking the row that leads to the best outcome within the fixed column).

\begin{figure}
\begin{mdframed}
\includegraphics[width=\textwidth]{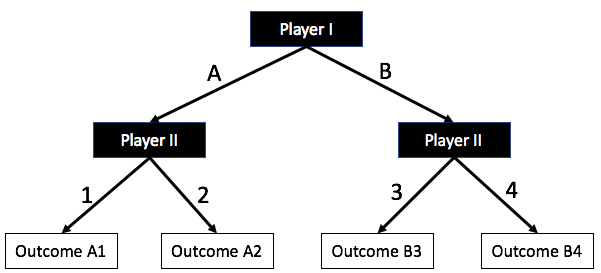}
\caption{A game in extensive form with two players playing one after the other. The second player's possible decisions depend on the first player's decision.} 
\label{fig-extensive-form}
\end{mdframed}
\end{figure}

[3] The Nash equilibrium is also defined for games in extensive form, for example, as its Subgame Perfect Equilibrium refinement. In a game in extensive form, as shown on Figure \ref{fig-extensive-form}, moves are causally dependent from each other: an equilibrium must be a consistent path from the root to a leaf and a player can only actually play at a node if that node is reached. The Subgame Perfect Equilibrium is obtained by a backward induction on the extensive form.

\begin{figure}
\begin{mdframed}
\includegraphics[width=\textwidth]{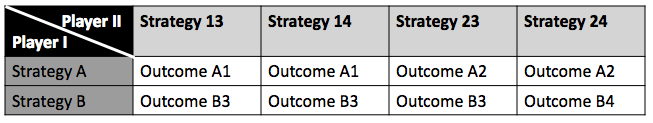}
\caption{The game from Figure \ref{fig-extensive-form}, expressed in normal form. The strategies of Player II are obtained with the cartesian product $\{ 1, 2 \} \times \{ 3, 4 \}$ } 
\label{fig-convert}
\end{mdframed}
\end{figure}

[4] The extensive form can be converted to a normal form by building strategies mapping each node in the tree to a decision, as shown on Figure \ref{fig-convert}. The rationale for this conversion is that players decide their strategy ahead of the game, with a plan of action for each node they may be playing at. This takes time out of the picture, as well as causal dependency, by considering all possible futures in advance. It works because causality only works forward in time. The Subgame Perfect Equilibrium for the extensive form then qualifies as a Nash equilibrium for the normal form. Here too, it becomes apparent on the normal form that players select a strategy independently from each other and hold the opponent's strategy for fixed.

[5] The idea that players act independently of each other in the Nash paradigm translates, in the extensive form and with perfect information, to the fact that a player, at any time in the game, can consider the past of the game, that is, the path leading to the current node, as given and fixed. This assumption appears to many as obvious, because the decision made at a node cannot causally impact what was made in the past, and thus must be independent from that past.

[6] Causality, though, is not the only source of dependency between the actions of the agents. Decisions can also be counterfactually correlated to each other. The general idea of a correlation is well accepted by the game theory community. In Douglas Hofstadter's super-rationality \cite{hofstadter} solution to the prisoner's dilemma, or any symmetric game in normal form, the two players' decisions are perfectly correlated---according to the definition of super-rationality itself---leading to a payoff optimization over the diagonal cells. While not mainstream, this approach appears as a credible proposal in the known landscape of solution concepts for games in normal form.

[7] Counterfactual correlation, also known as statistical correlation, does not know the arrow of time and is fully symmetric. In the prisoner's dilemma setting, the decisions are dependent yet potentially simultaneous, excluding any causality from the picture---relativity physicists would say that they are space-like-separated---making the distinction between causal dependency and counterfactual dependency completely apparent.

[8] In the same way, counterfactual correlation can also exist between the present and the past. While this is counterintuitive to accept, this follows from its symmetric nature: saying that the present is correlated with the past is equivalent to saying that the past is correlated with the present. The consequence of this remark is that the third assumption made in RCT must be seen as an axiom and not as an absolute truth. Assuming correlation rather than independence between the decisions of agents, even across time, neither contradicts the laws of physics, nor the other two axioms. In particular, the assumption or belief that there are correlations in the way rational agents act is not incompatible with the fundamental concepts behind rationality, and neither is it incompatible with free will.

[9] Dupuy \cite{dupuy} argued that a clear distinction between these two kinds of dependencies, causal and counterfactual, provides a solid basis for building an Alternate Rational Choice Theory (A-RCT). It can be obtained by dropping the third assumption presented above, and by replacing it with a full correlation between (i) the players' actions and (ii) the prediction of their actions. The set of fundamental assumptions then becomes:

\begin{enumerate}
\item People have rational preferences among outcomes.
\item Individuals maximize utility and firms maximize profits.
\item People assume that their actions are logically predictable, to the extent that these actions are based on full and relevant information, and that their decisions are thus correlated with past actions of others.
\end{enumerate}

[10] This alternate third assumption models players that (think that they) have very good predictive skills---Dupuy calls them essential predictors---in the sense that their prediction is not only right in the actual world, but also in all other possible worlds. This leads to a completely transparent theoretical setting in which players know as much as an omniscient external observer. 

\section{Counterfactual dependency across time}

[11] As previously said, two decisions separated by time could also be correlated to each other: if B is in the future of A, B could be counterfactually dependent on A. However, because of the symmetric nature of correlation, this is identical to saying that A, lying in the past of B, is counterfactually dependent on B. This is counterintuitive, albeit consistent with the laws of physics.

[12] Thus, the idea of correlated decisions can be stretched to extensive form games. We (Dupuy \cite{dupuy2}, Fourny et al \cite{fourny}) proposed a solution concept for games in extensive form called Perfect Prediction Equilibrium (PPE) in which the agents' decisions are assumed to be correlated to each other across time, providing an alternate account for the rationality of the players, in which they do not act independently (alternate third assumption). It is defined on games with perfect information, strict preferences, played by rational players and as such is an alternative to the Subgame Perfect Equilibrium (SPE), the latter corresponding to the equilibrium reached by agents rational in the RCT sense (RCT agents).

[13] In the Perfect Prediction Equilibrium, a rational agent in the alternate sense (A-RCT) considers that the other players, in the past, perfectly anticipated her decision, that their anticipation is thus correlated to her decision, and thus that past moves are likewise correlated to her decision. A rational player in the A-RCT sense (A-RCT agent) playing at a node N will not play certain moves because the decision to play them would be correlated to a deviation, in the past, from the path leading to N, thus leading to a logical contradiction. An A-RCT agent optimizes her utility across consistent paths (fix-points).

[14] An RCT agent, however, does not exclude such nodes and optimizes his utility across all available moves at any node, anticipating on each possible future.

[15] The fact that an A-RCT agent forbids herself to play certain moves because they are correlated with a different (inconsistent) path is often misunderstood. The remainder of this note aims at clarifying the most often raised points of criticism, advocating for the PPE as a reasonable Alternate Rational Choice Theory for games in extensive forms.

[16] An A-RCT agent has preferences in the same way as an RCT-agent, transitive and complete (1st assumption of RCT). These preferences are expressed with the same framework: ordinal payoffs with high preference for higher payoffs. An A-RCT agent optimizes her utility in the same way as an RCT-agent: facing the choice between several possible actions, both select the one with the highest utility (2nd assumption of RCT).

[17] A-RCT agents and RCT agents diverge in their behavior regarding their assessment of dependencies between decisions. While an RCT agent considers his decision to be independent from other player's decisions, an A-RCT agent will instead consider that her predictability allows other players to anticipate her decision, and that the past actions of the latter are correlated to her decision. In some settings, an A-RCT agent will exclude decisions that cannot causally follow from their anticipation from her payoff optimization domain, but this does not contradict the first two assumptions shared with RCT.

\section{Descriptive and normative support: falsifiable assumptions}

[18] Any candidate solution concept in game theory, but most generally science, must be formulated in a way that it can be contradicted by experience. The PPE predicts a game outcome on a tree satisfying the hypotheses. This outcome may or may not be identical to the one predicted by the SPE. Most importantly this outcome may or not happen in games played by real players, with reality being the final judge.

\begin{figure}
\begin{mdframed}
\includegraphics[width=\textwidth]{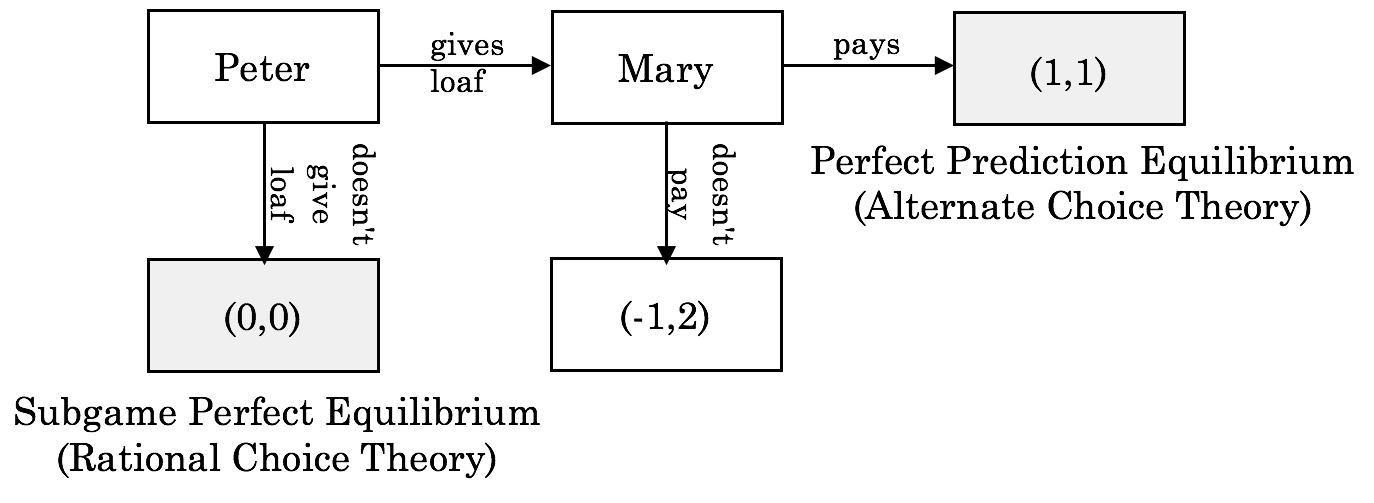}
\caption{A game that illustrates the difference between the two types of rationalities. Peter, the baker, can give a loaf of bread to Mary, the customer. If he does, Mary can choose to pay or not. In Rational Choice Theory (no correlation between their choices), the exchange does not take place because of backward induction. In  Alternate Rational Choice Theory, the asynchronous economic exchange takes place: Mary believes that there is a correlation, so that her not paying would logically (not causally) imply Peter's not having given her the loaf in the first place.} 
\label{figure1}
\end{mdframed}
\end{figure}

[19] There are some games in extensive form on which the PPE correctly predicts the behavior of some actual players and the SPE does not. A notable example is the asynchronous exchange, depicted on Figure \ref{figure1}. The payoffs correspond to Peter's, then Mary's. They have an ordinal, not cardinal, meaning and represent their relative preferences for the game outcomes.

[20] The game comes down to Mary going to the baker's (Peter). Peter may or may not give Mary a loaf of bread. If he does, Mary may or may not pay. The SPE predicts that Peter will not give any loaf to Mary, because he predicts that she would not pay: not paying maximizes Mary's utility (2 vs. 1) if she considers that Peter's giving her the loaf of bread is a fixed fact.

[21] However, in real life, many people behave honestly and pay. One could argue that law enforcement should be part of the picture in the game, yet there are many examples of people behaving honestly with no sword of Damocles hovering over them. This behavior is captured by the PPE, which correctly predicts that Peter hands over the loaf of bread and that Mary pays. Indeed, If Mary is an A-RCT agent, she considers that her not paying is correlated to Peter's not giving her the loaf of bread. She thus considers that no consistent timeline can exists in which she does not pay, and that it is hence rational for her to pay. Peter knows that Mary is an A-RCT agent, anticipates Mary's reasoning -- that she will pay -- and gives her the loaf of bread.

[22] This Alternate Rational Choice Theory has thus descriptive support. However, a feature of the PPE is that the equilibrium is always Pareto-optimal, which also brings in some normative support: A-RCT agents, to the extent that Pareto optimality is considered a good thing for the distribution of goods, have this additional benefit. Mongin \cite{mongin} contributed work on the notion of extended preferences, where players can optimize not only their preferences, but also optimize who they want to be, given the preferences of each person they can be. This could be extended to a choice that is also based on the kind of rationality. Seen from this point of view, if one were to choose between being behaving according to RCT or A-RCT, Pareto optimality weighs in the balance.

[23] Another example where A-RCT agents are better off than RCT agents---in the sense of the extended preference framework mentioned above---is Newcomb's problem \cite{gardner}. In Newcomb's problem, RCT agents are two-boxers because they consider that the predictor's anticipation is independent from their decision. A-RCT agents, on the other hand, are one-boxers, because they consider the prediction to be perfectly correlated with their decision. Newcomb's problem is hence an example where A-RCT agents get more (\$1,000,000) and RCT agents less (\$1,000). Although this is not always the case, still, A-RCT agents playing together will never reach a Pareto-dominated outcome.

\section{Two complementary theories of choice}

[24] A common criticism made to the PPE is that players seem to behave selflessly and are kind to each other, because their never select moves that do not ``please'' past players---the latter would have deviated otherwise. However, the PPE is a non-cooperative game theory paradigm. Players selfishly maximize their payoffs, but have a different view on the counterfactual dependencies between their actions and their anticipations, which leads to a different maximum than the SPE.

[25] Whether or not anticipations are correlated with future moves is orthogonal to being rational or not. This is comparable to Euclid's axiom: one can take it and build Euclidian geometry, or take others and obtain non-Euclidian geometries. Here, one can assume the past to be counterfactually independent and build the Nash equilibrium, or assume that it is correlated, and build non-Nashian equilibria.

[26] Is there room for an Alternate Rational Choice Theory? Newcomb's paradox points out that some people reason according to the Nash paradigm and pick two boxes, but other people pick one box because they model anticipations differently. It is a concrete experiment that often leads to polarizing views in the population, and demonstrates that agents do have at least two kinds of behavior. Both behaviors can be accounted for with preference functions and utility maximization, with a difference lying only in the assumptions made on counterfactual correlations. The behavior of A-RCT agents, as irrational as it appears to RCT agents -- because it breaks the third assumption -- can be modelled, predicted, accounted for and experimentally tested in the same way as RCT. In a comment to a presentation by John Dupr\'e about RCT back in 2009, David Kreps \cite{kreps} supported the idea that there may be other theories of Rational Choice as follows:

\begin{quote}
\begin{mdframed}
``When one says ``rational choice theory'', it sounds as if only one theory or model of choice could qualify. How could two distinct
theories or models both be rational? But people behave in different ways, depending
on the specific context and the more general social situation, and I
see no reason to privilege one universal model of behavior with the adjective
\emph{rational}.''

---David Kreps, 2009, Comments on Dupr\'e, Dupuy, and Bender
\end{mdframed}
\end{quote}

[27] This note aims by no means at advocating A-RCT over RCT, and even less at implying that RCT should be replaced. The thoroughly researched Nash equilibria cover the ground of agents considering their decisions to be uncorrelated with past actions, while we argue here that there is also a large, mostly unexplored territory with different, just as reasonable assumptions. This territory has both descriptive and normative potential and deserves a closer look.

\bibliographystyle{alpha}

\end{document}